\documentclass[12pt,preprint]{aastex}
\usepackage{natbib}
\usepackage{amsmath}
\usepackage{rotating}

\begin{document}

\title{Gamma-ray burst afterglow scaling relations for the full blast wave evolution}
\author{Hendrik J. van Eerten, Andrew I. MacFadyen}
\affil{
  Center for Cosmology and Particle Physics, Physics
  Department, New York University, New York, NY 10003}

\begin{abstract}
We demonstrate that gamma-ray burst afterglow spectra and light curves can be calculated for arbitrary explosion and radiation parameters by scaling the peak flux and the critical frequencies connecting different spectral regimes. Only one baseline calculation needs to be done for each jet opening angle and observer angle. These calculations are done numerically using high-resolution relativistic hydrodynamical afterglow blast wave simulations which include the two-dimensional dynamical features of expanding and decelerating afterglow blast waves. Any light curve can then be generated by applying scaling relations to the baseline calculations. As a result, it is now possible to fully fit for the shape of the jet break, e.g. at early time X-ray and optical frequencies. In addition, late-time radio calorimetry can be improved since the general shape of the transition into the Sedov-Taylor regime is now known for arbitrary explosion parameters so the exact moment when the Sedov-Taylor asymptote is reached in the light curve is no longer relevant. When calculating the baselines, we find that the synchrotron critical frequency $\nu_m$ and the cooling break frequency $\nu_c$ are strongly affected by the jet break. The $\nu_m$ temporal slope quickly drops to the steep late time Sedov-Taylor slope, while the cooling break $\nu_c$ first steepens then rises to meet the level of its shallow late time asymptote.
\end{abstract}

\section{Introduction}

Gamma-ray bursts (GRBs) are currently thought to result from the collapse of a massive star \citep{Woosley1993, MacFadyen1999} or a black hole-neutron star or neutron star-neutron star merger (e.g \citealt{Eichler1989, Paczynski1991}). During these processes, a collimated relativistic blast wave is launched into the circumburst medium. The emission from the blast wave is commonly referred to as the \emph{afterglow} of the burst and can be observed throughout the broadband spectrum as the blast wave decelerates and radiates at progressively longer wavelengths \citep{Meszaros1997}. Ever since the discovery of the first afterglows, these have been modeled succesfully by combining a model for the blast wave dynamics with a sychrotron radiation model, where shock-accelerated particles radiate by interacting with a shock-generated magnetic field (e.g. \citealt{Wijers1997, Wijers1999, Frail2000, PK2002}). Analytically tractable solutions for the dynamics are the self-similar Blandford-McKee (BM, \citealt{Blandford1976}) and Sedov-von Neumann-Taylor (ST, \citealt{Sedov1959, Taylor1950, vonNeumann1961}) solutions describing, respectively, the ultra-relativistic and non-relativistic phase of the blast wave evolution. At early time lateral spreading of the collimated jet has not yet set in and the outflow is purely radial, while at late times the jet will have become truly spherical, allowing the application of spherically symmetric solutions in both cases. As of yet, no analytical solution exists that fully captures the intermediate stage of the blast wave evolution, where the blast wave becomes transrelativistic, inhomogenous along the shock front \citep{Zhang2009, vanEerten2011jetspreading} and decollimates. Early attempts assumed a homogeneous shock front \citep{Rhoads1999} or spherical outflow \citep{Huang1999}, while even recently studies (e.g. \citealt{Granot2011}) do not account for the radial structure of the jet.

The practical implications of the fact that the blast wave evolution is determined by a very small number of variables such that scalings between different explosion energies and circumburst density are possible were not fully realized until very recently. The scalings apply in the asymptotic self-similar limits, but also in the intermediate regime where the two-dimensional nature of jet decollimation is in full effect. This made it possible to use only a small set of simulations for different initial jet opening angles as a basis for a simulation-based fit code that can be applied to broadband afterglow data \citep{vanEerten2011boxfit}.

But even though a complete recalculation of the dynamics of the blast wave is no longer necessary, there remained the equations of radiative transfer of (a representative number of) rays through the evolving jet, that have to be solved for each datapoint fit iteration. For a large number of iterations and datapoints this procedure remains computationally expensive and requires the use of a parallel computer. 

The current study shows that the calculation time for a given light curve or spectrum can be further reduced. We demonstrate that scalability between blast waves has straightforward implications for scalability between light curves. In \S \ref{scalings_section} we describe how the scaling relations for the dynamics of blast waves can be used to scale between light curves as well. We show that the scalings remain unchanged between the BM and ST regimes, and in \S \ref{numerics_section} we demonstrate numerically that the scalings also hold in the intermediate regime and for off-axis observers. We discuss our findings in \S \ref{discussion_section}.

\section{Full evolution scaling relations}
\label{scalings_section}

We take the BM solution for the impulsive injection of an energy $E_{iso}$ in a homogeneous medium with density $\rho_0 = n_0 m_p$ (with $n_0$ the number density and $m_p$ the proton mass) as the intitial condition. In this paper we will discuss only adiabatic blast waves exploding in a homogeneous interstellar medium, but all conclusions drawn here can be generalized to a stellar wind environment as well, which will be presented in a forthcoming publication. Only a small number of independent dimensionless combinations of the variables determining the fluid state at distance $r$, angle $\theta$ and source frame time $t_e$ exists: $A \equiv r / ct_e$ (with $c$ the speed of light), $B \equiv E_{iso} t_e^2 / \rho_0 r^5$, $\theta$ and $\theta_0$ (the initial jet half-opening angle). Any fluid quantity can be expressed as a dimensionless combination (e.g. $n / n_0$ for the local number density $n$) and therefore as a function of the dimensionless variables. It follows that dimensionless fluid quantities are invariant under scalings that leave the dimensionless variables invariant. \cite{vanEerten2011boxfit} made practical use of the scalings

\begin{eqnarray}
 E'_{iso} & = & \kappa E_{iso}, \nonumber \\
 \rho_0' & = & \lambda \rho_0, \nonumber \\
 r' & = & (\kappa / \lambda)^{1/3} r, \nonumber \\
 t_e' & = & (\kappa / \lambda)^{1/3} t_e.
\end{eqnarray}

\begin{figure}
 \centering
  \includegraphics[width=0.6\columnwidth]{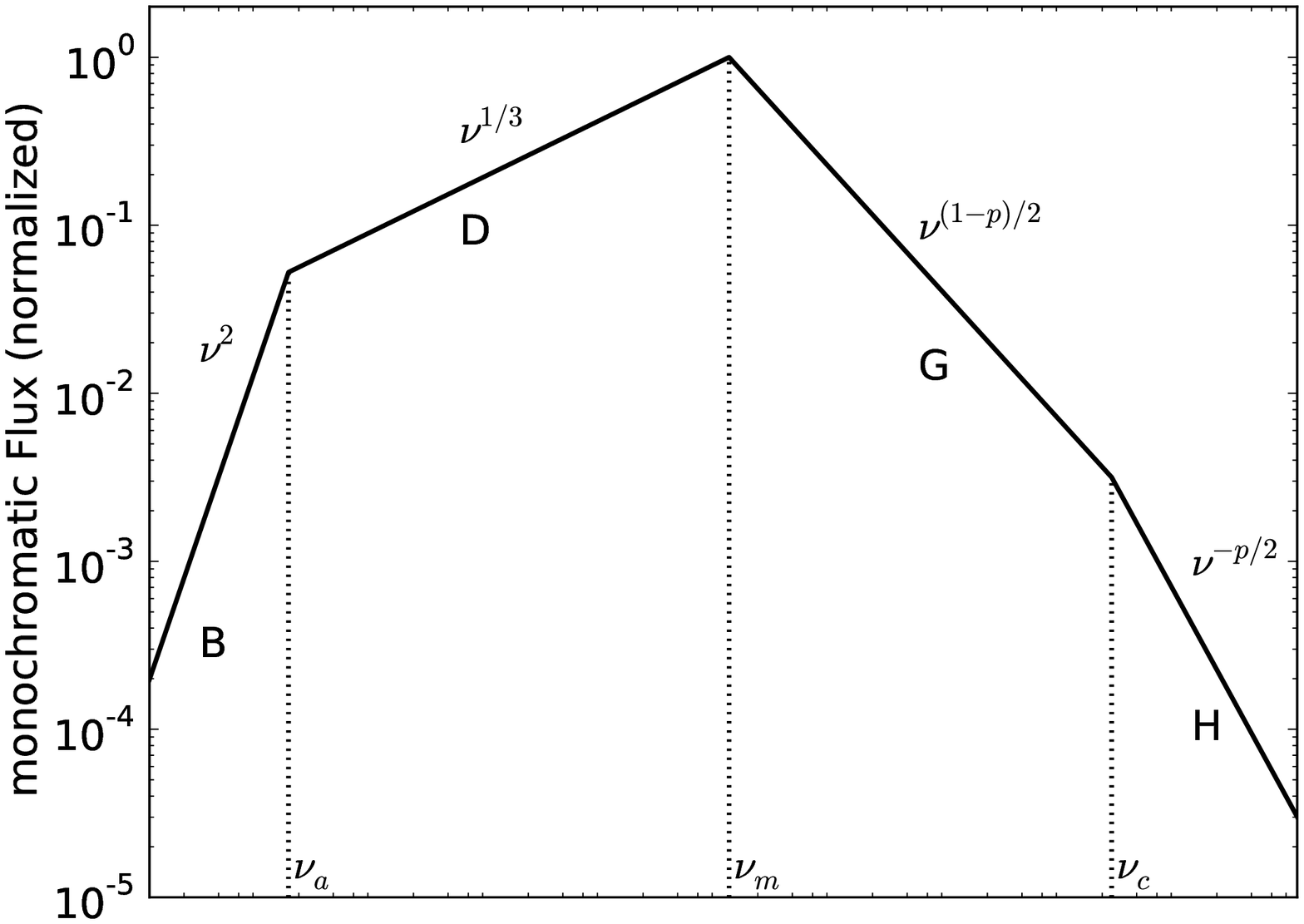}
 \includegraphics[width=0.6\columnwidth]{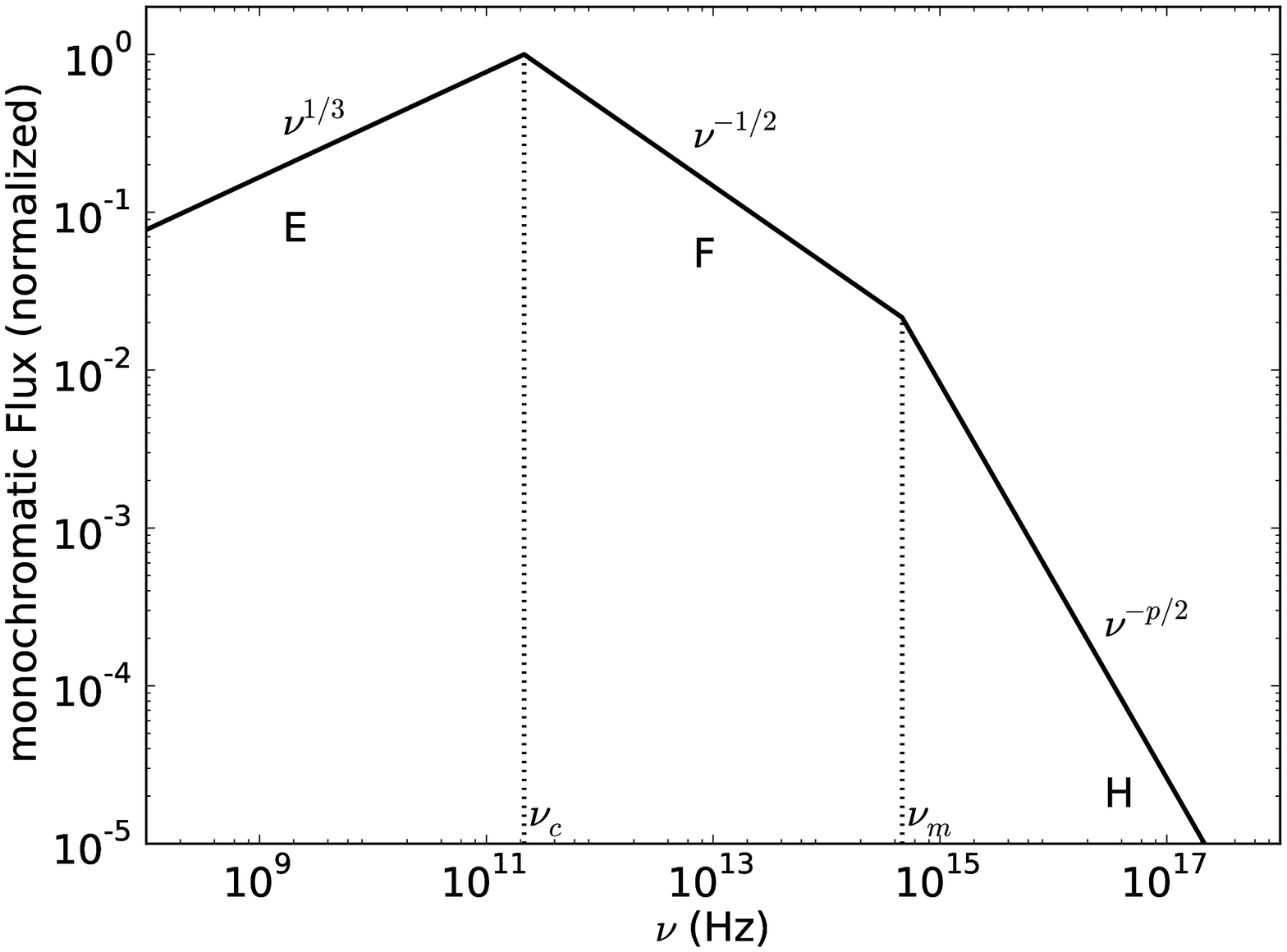}
 \caption{Synchrotron spectra with different orderings of the critical frequencies. The different power law segments are labeled by letters (following \citealt{Granot2002}) and the spectral slopes are indicated in the plots.}
 \label{powerlaws_figure}
\end{figure}

The synchrotron emission spectra from the expanding blast wave can be described locally by a series of connecting power laws, see Fig. \ref{powerlaws_figure}. Below the critical frequency $\nu_{a}$ the medium is optically thick due to synchrotron self-absorption. Critical frequency $\nu_m$ marks the synchrotron break frequency. Above the cooling-break frequency $\nu_c$ the accelerated electrons lose their energy too quickly to radiate fully at these frequencies. In detailed numerical models the evolution of the electron distribution is traced explicitly as it advects into the non-radiating, slow-moving and dilute downstream region, see e.g. \cite{Downes2002, vanEerten2010transrelativistic}. Alternatively, a steady state for the radiating fluid is assumed and the global cooling time $t_c$ is equated to the duration of the explosion $t_e$ \citep{Sari1998}. Both approaches lead to qualitatively the same behavior \citep{vanEerten2010offaxis}.

Synchrotron radiation from shock-accelerated electrons in a shock-generated magnetic field is parametrized as follows: $p$ denotes the power law slope of the shock-accelerated electron distribution, $\epsilon_B$ the fraction of magnetic energy relative to thermal energy, $\epsilon_e$ the fraction of downstream thermal energy density in the accelerated electrons, and $\xi_N$ the fraction of the downstream particle number density that participates in the shock-acceleration process. Typical values for these parameters are $p \sim 2.5$, $\epsilon_B \sim 0.01$, $\epsilon_e \sim 0.1$, $\xi_N \sim 1$. These parameters can be constrained from broadband afterglow data. At late times, maintaining $\xi_N \sim 1$ sometimes leads to unphysically low values for the lower cut-off Lorentz factor of the accelerated particle distribution (which determines $\nu_m$). For a discussion, see e.g. \cite{vanEerten2010transrelativistic}, which explores (scale-invariant) evolution of $\xi_N$.

The synchrotron emission coefficient $j_\nu$ for a local distribution of particles follows $j_\nu \propto \xi_N n B f(\nu, \nu_m, \nu_c) / \gamma^2 (1- \beta \mu)^2$ in the source frame, where number density $n$ and magnetic field $B$ are in the frame comoving with the fluid, $\gamma$ denotes the fluid Lorentz factor, $\beta$ the fluid velocity divided by $c$ and $\mu$ the cosine of the angle between the fluid velocity and the observer direction. The function $f(\nu, \nu_m, \nu_c)$ is given by the synchrotron spectrum as shown in Fig \ref{powerlaws_figure} (without self-absorption, which is included in the absorption coefficient when enabled) and is normalized to one where the spectrum peaks. In the optically thin case, the observed flux is given by
\begin{equation}
F = \frac{1+z}{d_L^2} \int \mathrm{d}V j_\nu / \gamma (1 - \beta \mu).
\label{flux_equation}
\end{equation}
Here $d_L$ is the luminosity distance and $z$ the redshift. Although self-absorption is shown in Fig \ref{powerlaws_figure} for the case where $\nu_a < \nu_m < \nu_c$, we will concentrate on the case where observer frequency $\nu > \nu_{a}$ and leave a detailed treatment of self-absorption to future work. The leading order dependencies of the observed flux on the model parameters can be calculated from eq. \ref{flux_equation}, the shape of the synchrotron spectrum and a model for the dynamics that sets the radius of the afterglow blast wave as well as the post-shock values of $n$ and $B$, $\nu_m$ and $\nu_c$ (the latter two via their dependence on the local fluid state, e.g. \citealt{Sari1998}). These are shown in table \ref{scalings_table}.

\begin{table}
\centering
\begin{tabular}{|l|l|l|l|}
\hline
$F$ or $\nu$ & leading order scalings & $\kappa$ & $\lambda$ \\
\hline
$F_{B,BM}$ & $(1+z) E_{iso}^{1/2} n_0^{-1/2} \epsilon_e^{1} \epsilon_B^{0} \xi_N^{-1} t^{1/2} \nu^{2}$ & $\kappa^{2/3}$ & $\lambda^{-2/3}$ \\
$F_{B,ST}$ & $(1+z) E_{iso}^{4/5} n_0^{-4/5} \epsilon_e^{1} \epsilon_B^{0} \xi_N^{-1} t^{-2/5} \nu^{2}$ &  &  \\
\hline
$F_{D,BM}$ & $(1+z) E_{iso}^{5/6} n_0^{1/2} \epsilon_e^{-2/3} \epsilon_B^{1/3} \xi_N^{5/3} t^{1/2} \nu^{1/3}$ & $\kappa^1$ & $\lambda^{1/3}$ \\
$F_{D,ST}$ & $(1+z) E_{iso}^{7/15} n_0^{13/15} \epsilon_e^{-2/3} \epsilon_B^{1/3} \xi_N^{5/3} t^{8/5} \nu^{1/3}$ & & \\ 
\hline
$F_{E,BM}$ & $(1+z) E_{iso}^{7/6} n_0^{5/6} \epsilon_e^{0} \epsilon_B^{1} \xi_N^{1} t^{1/6} \nu^{1/3}$ & $\kappa^{11/9}$ & $\lambda^{7/9}$ \\
$F_{E,ST}$ & $(1+z) E_{iso}^{1} n_0^{1} \epsilon_e^{0} \epsilon_B^{1} \xi_N^{1} t^{2/3} \nu^{1/3}$ & & \\
\hline
$F_{F,BM}$ & $(1+z) E_{iso}^{3/4} n_0^{0} \epsilon_e^{0} \epsilon_B^{-1/4} \xi_N^{1} t^{-1/4} \nu^{-1/2}$ & $\kappa^{2/3}$ & $\lambda^{1/12}$ \\
$F_{F,ST}$ & $(1+z) E_{iso}^{1/2} n_0^{1/4} \epsilon_e^{0} \epsilon_B^{-1/4} \xi_N^{1} t^{1/2} \nu^{-1/2}$ &  &  \\
\hline
$F_{G,BM}$ & $(1+z) E_{iso}^{(p+3)/4} n_0^{1/2} \epsilon_e^{p-1} \epsilon_B^{(1+p)/4} \xi_N^{2-p} t^{3(1-p)/4} \nu^{(1-p)/2}$ & $\kappa^1$ & $\lambda^{(1+p)/4}$ \\
$F_{G,ST}$ & $(1+z) E_{iso}^{(5p+3)/10} n_0^{(19-5p)/20} \epsilon_e^{p-1} \epsilon_B^{(1+p)/4} \xi_N^{2-p} t^{(21-15p)/10} \nu^{(1-p)/2}$ &  &  \\
\hline
$F_{H,BM}$ & $(1+z) E_{iso}^{(p+2)/4} n_0^{0} \epsilon_e^{p-1} \epsilon_B^{(p-2)/4} \xi_N^{2-p} t^{(2-3p)/4} \nu^{-p/2}$ & $\kappa^{2/3}$ & $\lambda^{(3p-2)/12}$ \\
$F_{H,ST}$ & $(1+z) E_{iso}^{(p)/2} n_0^{(2-p)/4} \epsilon_e^{p-1} \epsilon_B^{(p-2)/4} \xi_N^{2-p} t^{(4-3p)/2} \nu^{-p/2}$ &  &  \\
\hline
\hline
$F_{peak,BM}$ & $(1+z) E_{iso}^1 n_0^{1/2} \epsilon_e^{0} \epsilon_B^{1/2} \xi_N^1 t^{0}$ & $\kappa^1$ & $\lambda^{1/2}$ \\
$F_{peak,ST}$ & $(1+z) E_{iso}^{4/5} n_0^{7/10} \epsilon_e^{0} \epsilon_B^{1/2} \xi_N^1 t^{3/5}$ & & \\
\hline
$\nu_{m,BM}$ & $E_{iso}^{1/2} n_0^0 \epsilon_e^2 \epsilon_B^{1/2} \xi_N^{-2} t^{-3/2}$ & $\kappa^0$ & $\lambda^{1/2}$ \\
$\nu_{m,ST}$ & $E_{iso}^{1} n_0^{-1/2} \epsilon_e^2 \epsilon_B^{1/2} \xi_N^{-2} t^{-3}$ & & \\ 
\hline
$\nu_{c,BM}$ & $E_{iso}^{-1/2} n_0^{-1} \epsilon_e^0 \epsilon_B^{-3/2} \xi_N^0 t^{-1/2}$ & $\kappa^{-2/3}$ & $\lambda^{-5/6}$ \\
$\nu_{c,ST}$ & $E_{iso}^{-3/5} n_0^{-9/10} \epsilon_e^0 \epsilon_B^{-3/2} \xi_N^0 t^{-1/5}$ & & \\
\hline
$\nu_{a, BM}$ & $E_{iso}^{1/5} n_0^{3/5} \epsilon_e^{-1} \epsilon_B^{1/5} \xi_N^{8/5} t^0$ & $\kappa^{1/5}$ & $\lambda^{3/5}$ \\
$\nu_{a, ST}$ & $E_{iso}^{-1/5} n_0^{1} \epsilon_e^{-1} \epsilon_B^{1/5} \xi_N^{8/5} t^{6/5}$ &  &  \\
\hline
\end{tabular}
\caption{Scalings for flux in different spectral regimes, both in the relativistic BM limit and non-relativistic ST limit. Note that  $t$ and $\nu$ are expressed in the frame where redshift $z = 0$. In the observer frame, $\nu_\oplus = \nu / (1+z)$ and $t_\oplus = t (1+z)$. The $\kappa$ and $\lambda$ columns denote the corresponding scaling of the flux under a scaling of energy, time and density according to eq. \ref{lightcurve_scalings_equation}. The scaling of $\nu_a$ applies to the case where $\nu_a < \nu_m < \nu_c$.}
\label{scalings_table}
\end{table}

So far, no new arguments have been presented. \emph{However, eq. \ref{flux_equation} directly implies that scale invariance on the dynamical level leads to scale invariance for the flux within a given spectral regime} (e.g. when $j_\nu$ depends on $\nu_m$, $\nu_c$ and $\nu$ according to a single combination of power laws). The table shows this explicitly for the various spectral regimes  and the BM and ST limits. Additionally, we have expressed $t$ and $\nu$ at $z=0$ instead of the observer frame, which emphasizes that different redshift scalings of the fluxes between the different spectral regimes are a feature of the frame in which frequency and time are expressed. Light curves and spectra can be mapped onto different redshifts in a straightforward manner via a single $(1+z)$ dependency. Finally, also from eq. \ref{flux_equation}, the dependence on the radiation parameters $\epsilon_e$, $\epsilon_B$ and $\xi_N$ remains unchanged at all times.

The fact that observer time $t$ scales the same as $t_e$ and $r$ (that have been integrated over in order to obtain the flux) follows from the $t = t_e - \mu r / c$ constraint that matches different emission times to a single arrival time. In table \ref{scalings_table} we have tabulated the effect of applying the mapping
\begin{eqnarray}
 E'_{iso} & = & \kappa E_{iso}, \nonumber \\
 n_0' & = & \lambda n_0, \nonumber \\
 t' & = & (\kappa / \lambda)^{1/3} t,
\label{lightcurve_scalings_equation}
\end{eqnarray}
to the observed flux. This scaling remains unchanged between early time BM and late time ST for all spectral regimes, e.g. $F_{D,BM}' / F_{D,BM} = F_{D,ST}' / F_{D,ST} = \kappa \lambda^{1/3}$, independent of jet opening angle or observer angle.

\section{Numerical verification}
\label{numerics_section}

\begin{figure}
 \centering
  \includegraphics[width=0.51\columnwidth]{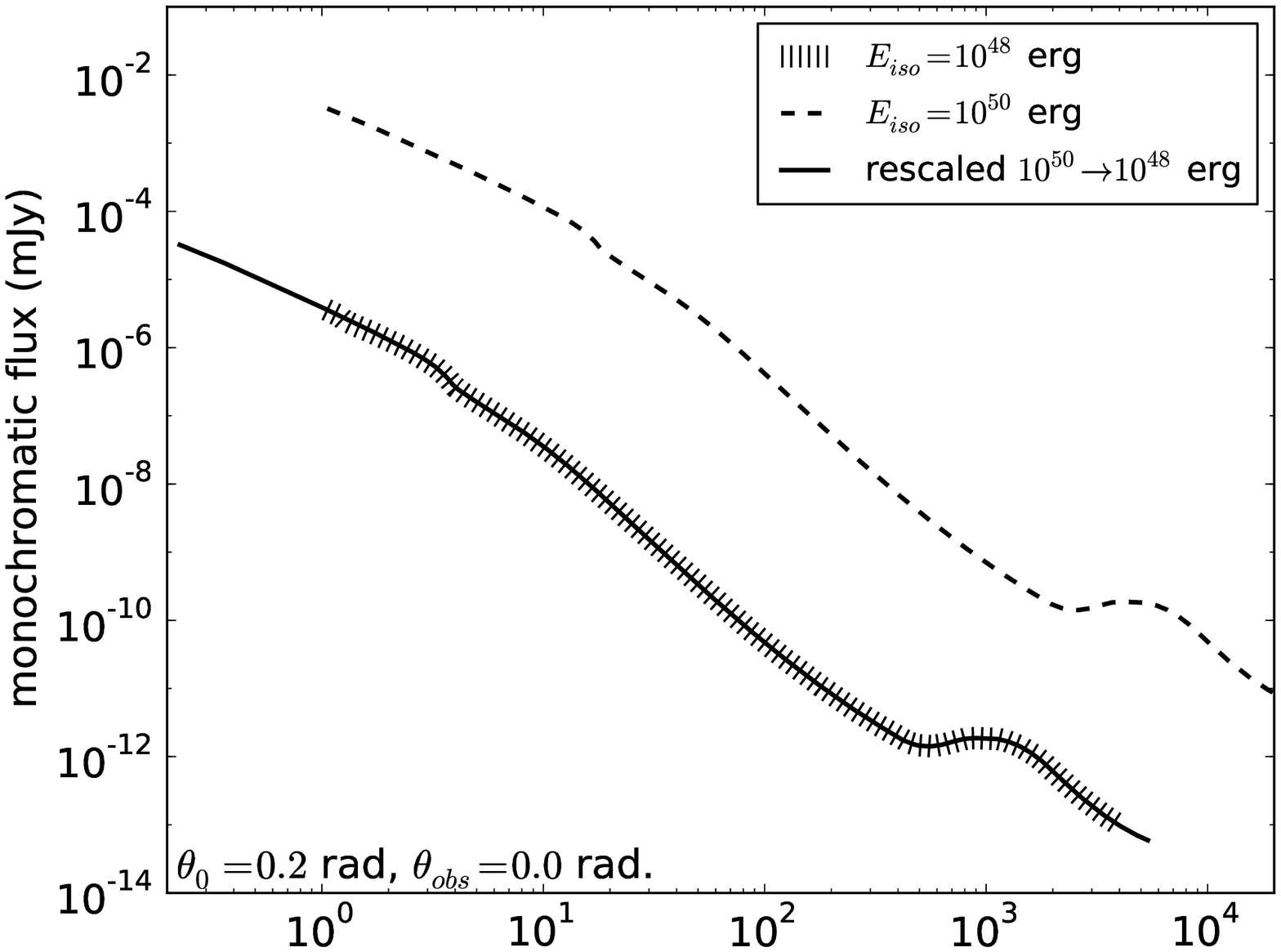}
 \includegraphics[width=0.51\columnwidth]{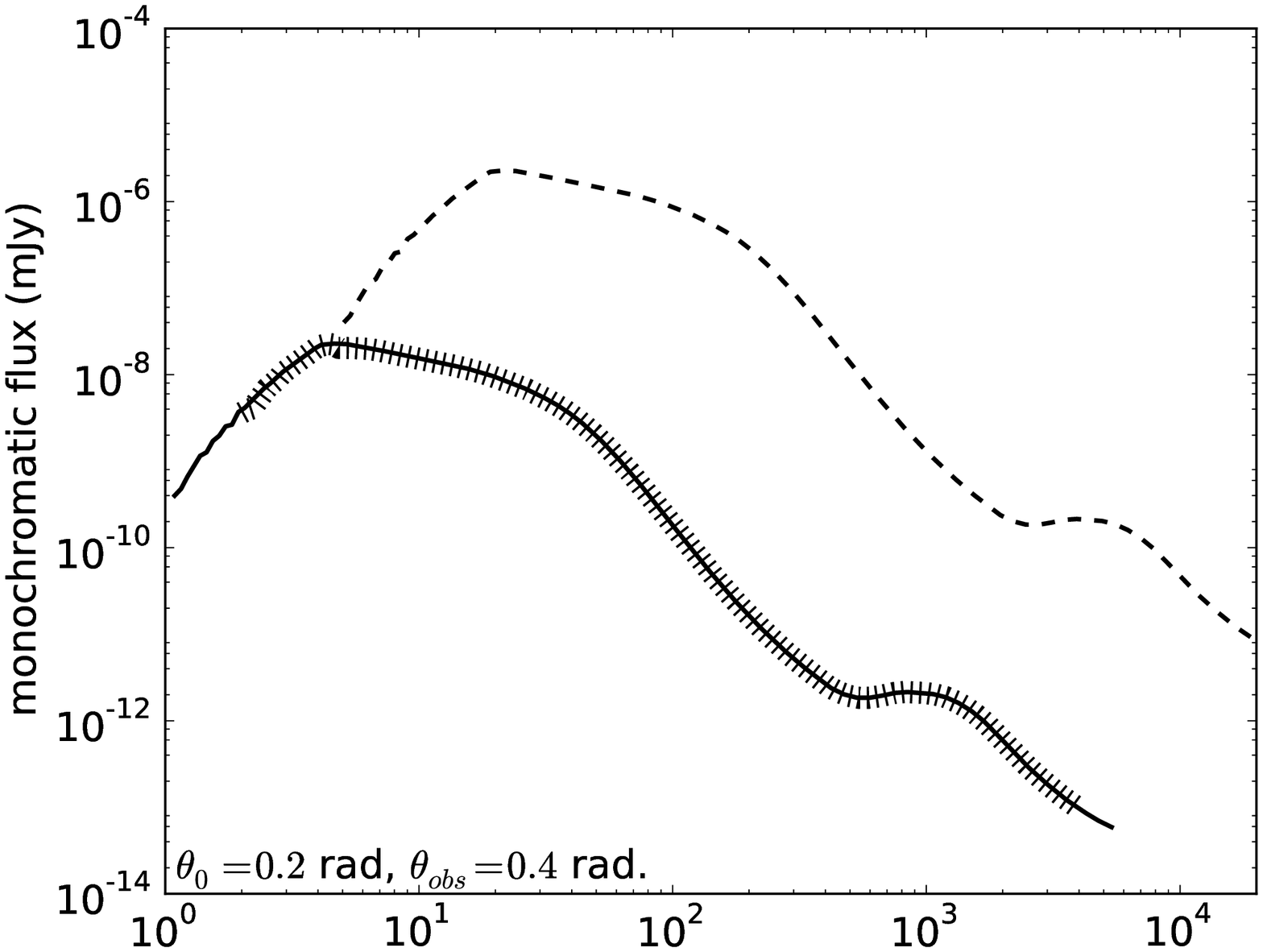}
 \includegraphics[width=0.51\columnwidth]{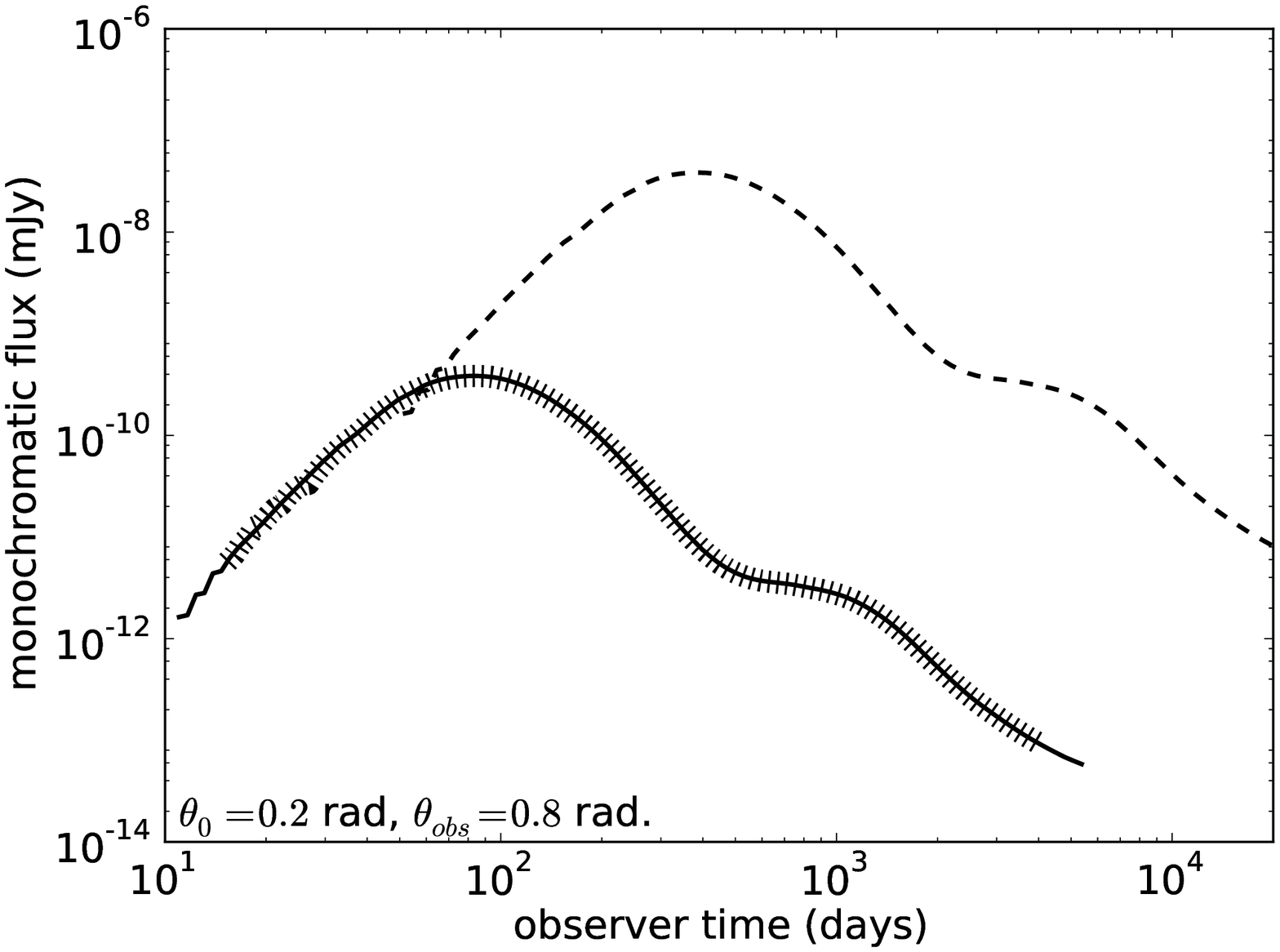}
 \caption{Demonstrations of scaling between optical light curves ($\nu = 4.56 \times 10^{14}$ Hz) with $E_{iso} = 10^{48}$ erg and $E_{iso} = 10^{50}$ erg, for different observer angles $\theta_{obs}$. Other parameters are set as follows: $n_0 = 10^{-3}$ cm$^{-3}$, $\theta_0 = 0.2$ rad, $p = 2.5$, $\epsilon_B = 0.01$, $\epsilon_e = 0.1$, $\xi_N = 1.0$, $z = 0$, $d_L = 10^{28}$ cm. The legend in the top plot refers to all plots.}
 \label{lightcurves_figure}
\end{figure}

Fig. \ref{lightcurves_figure} demonstrates that the scaling between energies applies at all times. The light curves remain in spectral regime G throughout their evolution, so $F'(E'_{iso}, t') = \kappa F(E_{iso}, t)$. All light curves were taken from the dataset used in \cite{vanEerten2011sgrb}, available from the on-line afterglow library at \url{http://cosmo.nyu.edu/afterglowlibrary}.

\begin{figure}
 \centering
  \includegraphics[width=0.51\columnwidth]{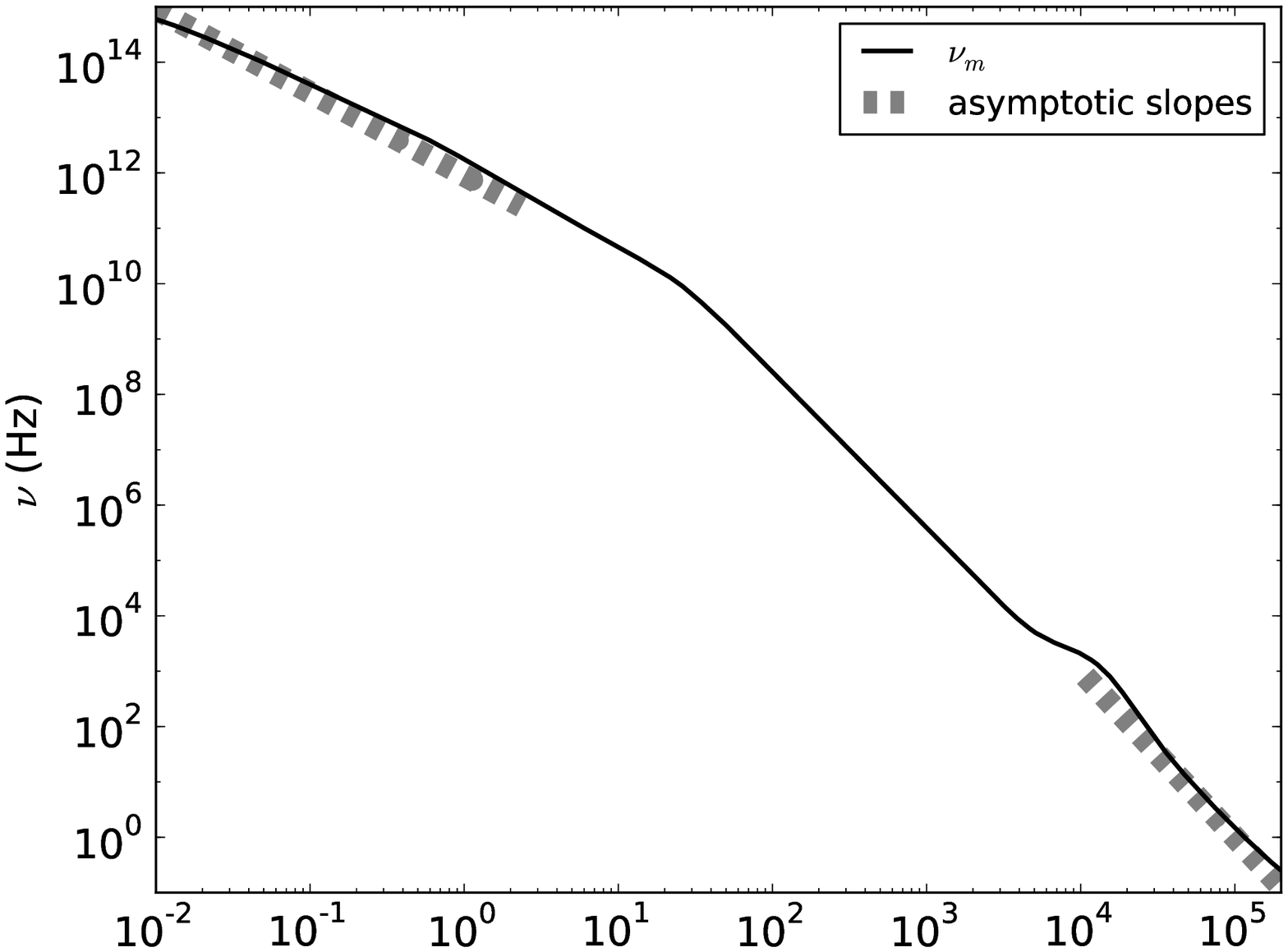}
 \includegraphics[width=0.51\columnwidth]{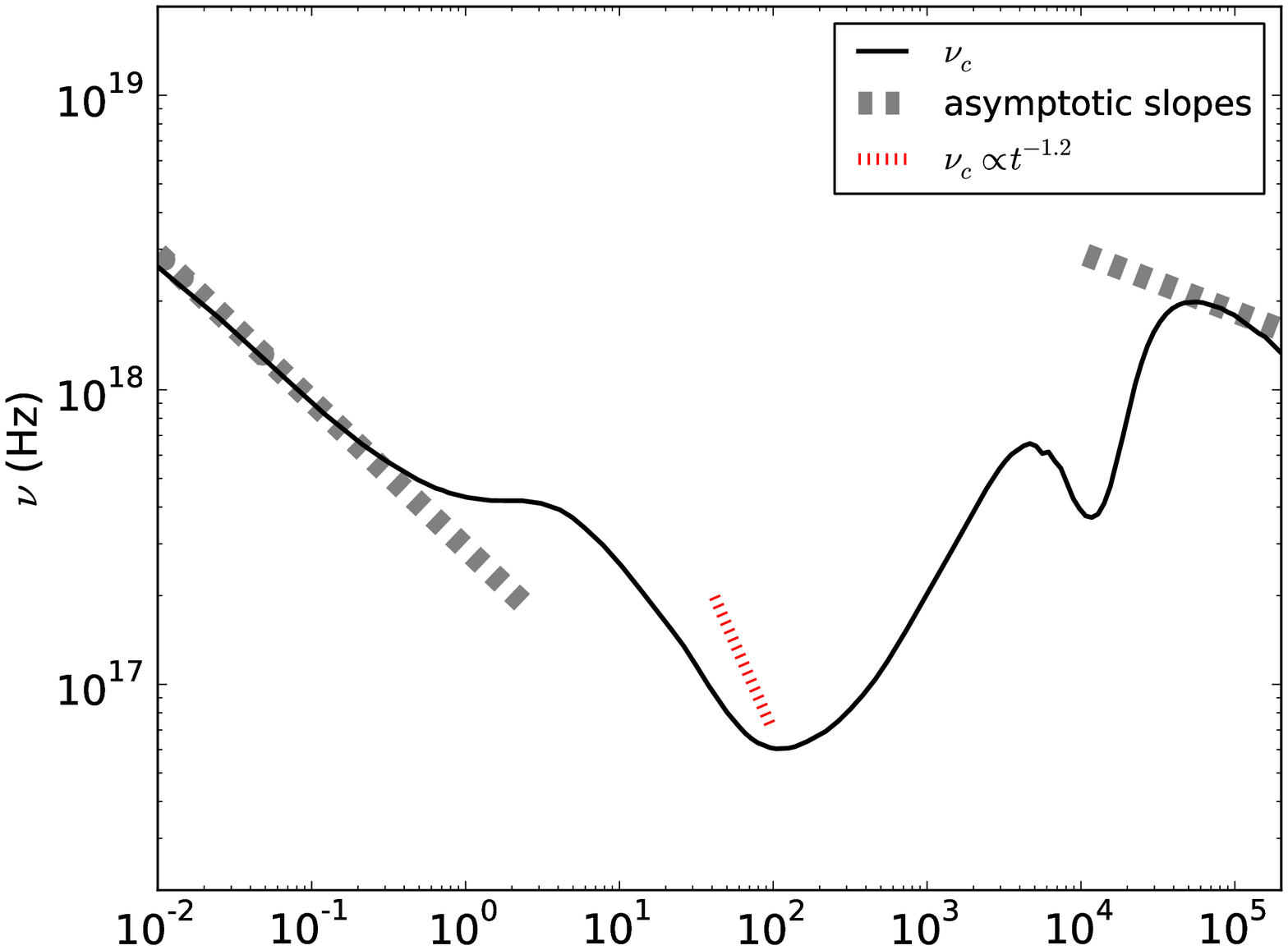}
 \includegraphics[width=0.51\columnwidth]{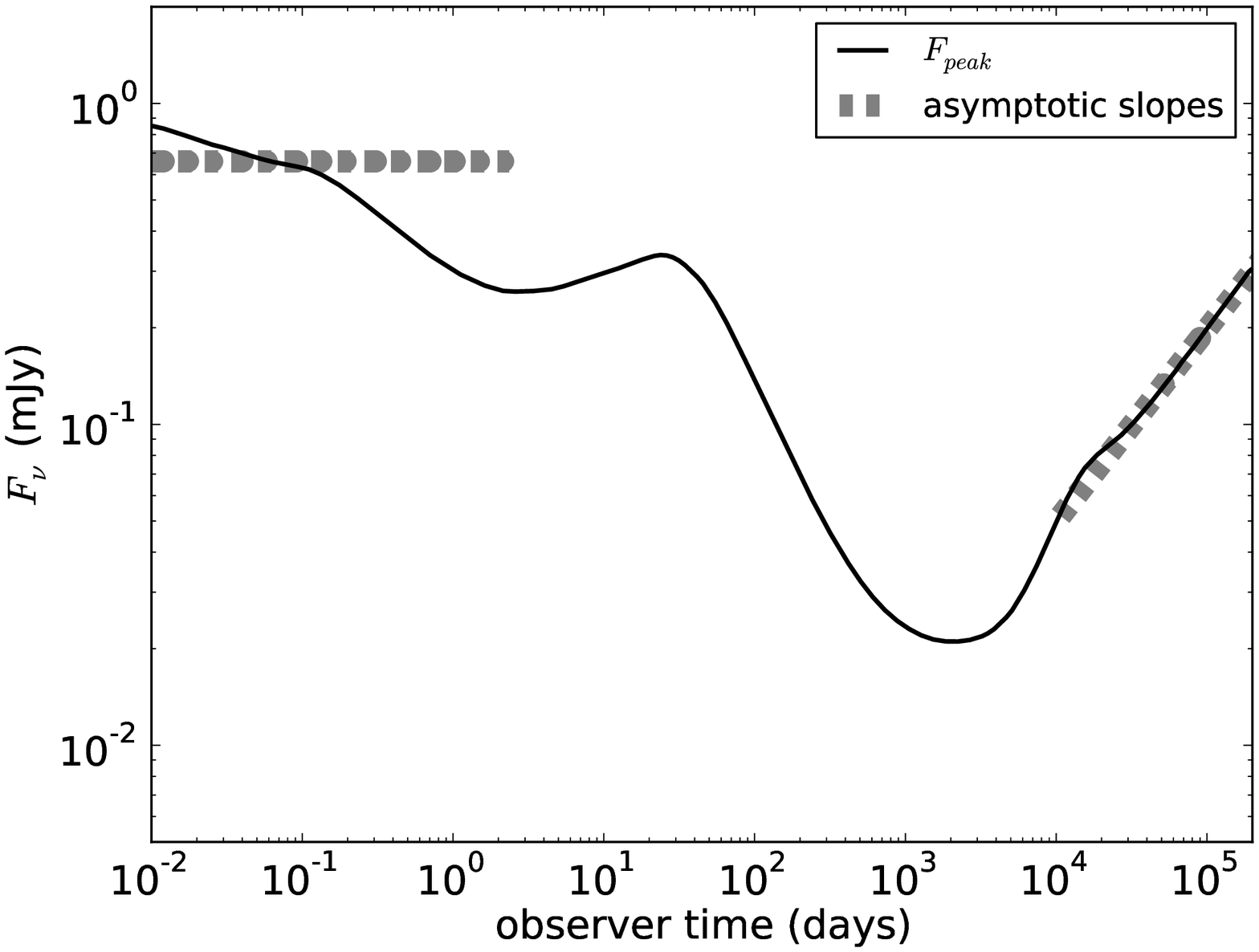}
 \caption{Evolution of critical quantities for on-axis observers. The following parameters were used: $E_{iso} = 5 \times 10^{52}$ erg, $n_0 = 10^{-3}$ cm$^{-3}$, $\theta_0 = 0.2$ rad, $p = 2.5$, $\epsilon_B = 0.01$, $\epsilon_e = 0.1$, $\xi_N = 1.0$, $z = 0$, $d_L = 10^{28}$ cm. From top to bottom, the plots show $\nu_m$, $\nu_c$ and $F_{peak}$. The BM and ST asymptotic slopes are indicated by dashed lines. The dotted line in the center of the $\nu_c$ plot indicates the temporal slope of $\nu_c$ found by \cite{Filgas2011} for GRB 091127.}
 \label{criticals_figure}
\end{figure}

To generate light curves that do not lie on the asymptote of a single spectral regime we need to take the spectral evolution into account. As shown in table \ref{scalings_table}, both the critical frequencies and peak fluxes obey the same scalings throughout the blast wave evolution. Therefore, once we know their time evolution for a given set of parameters ($E_{iso}$, $n_0$, $\theta_0$) we can use these as a baseline to generate the connected power-law spectra for arbitrary values of the jet parameters. In Fig. \ref{criticals_figure} we plot the time evolution of $F_{peak}, \nu_c, \nu_m$ for $\theta_0 = 0.2$ rad and an on-axis observer. They have been calculated using the methods presented in \cite{vanEerten2011boxfit} and are ultimately based on the simulations described therein. All simulations were performed using the \textsc{ram} adaptive-mesh refinement (AMR) relativistisc hydrodynamics (RHD) code \citep{Zhang2006}. For details, see \cite{vanEerten2011boxfit}.

At early times $\nu_c$ and $F_{peak}$ differ numerically from their asymptotically expected time evolution because the blast wave is initially underresolved in the RHD simulations. However, all explosion energy is included in the initial conditions of the simulations, and the resulting drop in blast wave Lorentz factor is temporary (for more details, see \citealt{vanEerten2011boxfit, Zhang2009}). But initially the observed flux level and critical frequencies are impacted due to the temporary decrease in beaming. However, the scalability of the jet dynamics justifies the computational cost of extremely high resolution blast wave simulations and these will be presented in future work.

For jet half-opening angle $\theta_0 = 0.2$, a jet break occurs around $\sim 10$ days. As a result, the time evolutions change significantly. The temporal slope of $\nu_m$ turns over quickly from $\nu_m \propto t^{-3/2}$ (BM) to the far steeper ST slope $\nu_m \propto t^{-3}$. The temporal slope for $\nu_c$ is not only less steep at late times, its late time ST asymptote also lies significantly higher than the early time BM asymptote, and as a result $\nu_c$ actually \emph{rises} for some time after the jet break. This effect will be less severe for larger opening angles, since $\nu_c \propto E_{iso}^{-3/5}$ in the BM regime and $\nu_c \propto E_j^{-3/5}$ in the ST regime, where $E_j$ the total energy in both jets (and therefore in the final ST sphere). The two energies are related via $E_j \approx E_{iso} \theta_0^2 / 2$. It is also worth noting that, before rising, the temporal slope of $\nu_c$ temporarily \emph{steepens} beyond $-1/2$. A steepening of the cooling break frequency to $\nu_c \propto t^{-1.2}$ has recently been observed in GRB 091127 by comparing optical and X-ray data \citep{Filgas2011}. Our plot shows that this is, in principle, not inconsistent with simulations (and therefore with the standard model, since we do not expand upon the standard synchrotron framework by including features like evolving microphysics parameters such as $\epsilon_B$). However, we caution against overinterpretation of the post-break $\nu_c$ evolution because our approach to electron cooling (based on \citealt{Sari1998}) relies on a single global cooling time approximation rather than on tracing the local accelerated electron distribution (for a comparison between the two approaches, see \citealt{vanEerten2010offaxis}. In the example there, $\nu_c$ for local cooling is typically higher by a factor $\sim 5$). Given this caveat for $\nu_c$, a clear steepening of $\nu_m$ and $\nu_c$ immediately post jet break is a general prediction of our study, with the steepening of $\nu_m$ being more robust. 

The final feature in all three evolution plots is the onset of the counterjet around $\sim 250$ days, resulting in a relative increase of $F_{peak}$ and $\nu_m$ and a decrease in $\nu_c$. This effect is strongest around $\sim 1500$ days.

\begin{figure}
 \centering
  \includegraphics[width=0.51\columnwidth]{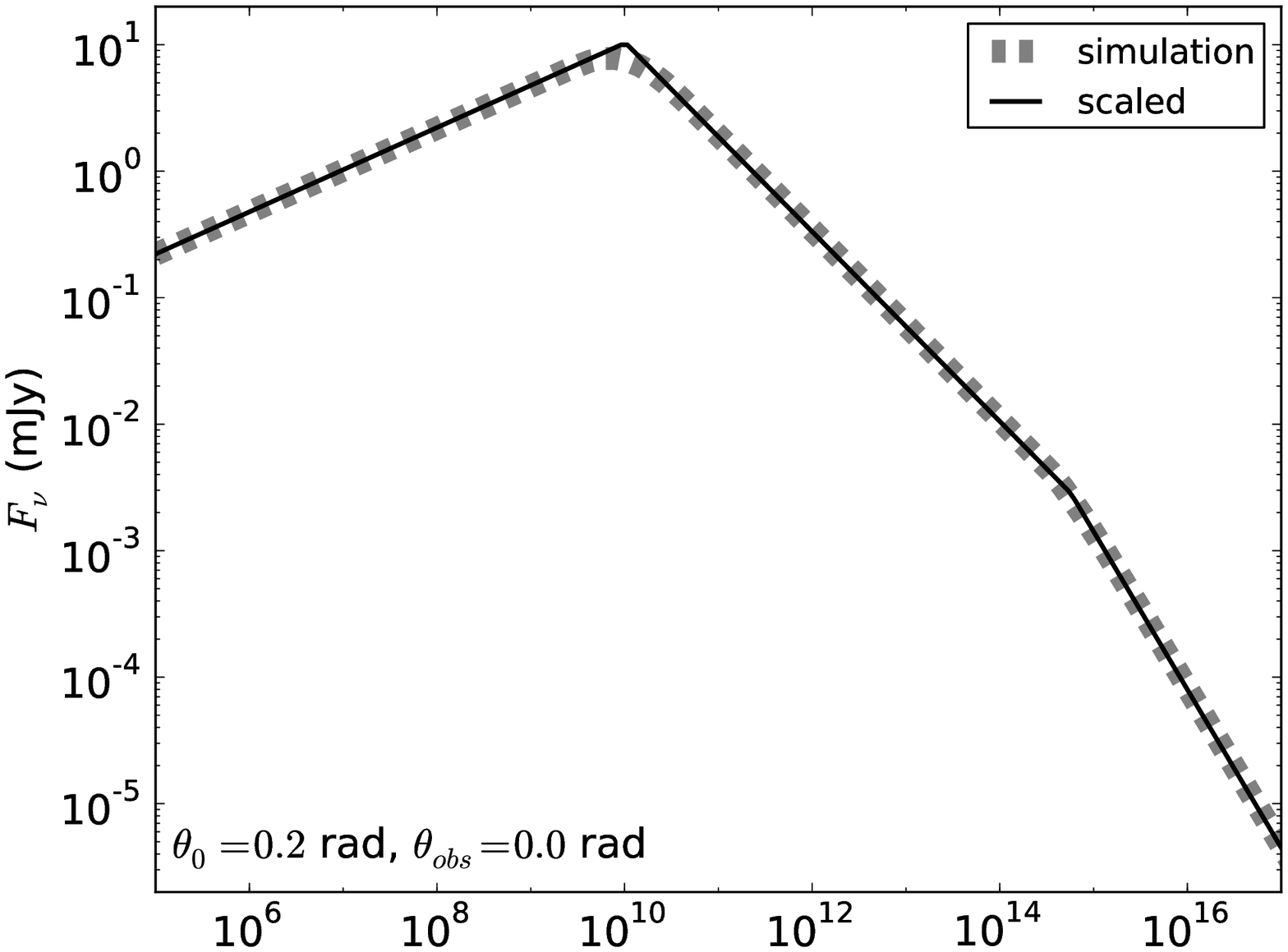}
  \includegraphics[width=0.51\columnwidth]{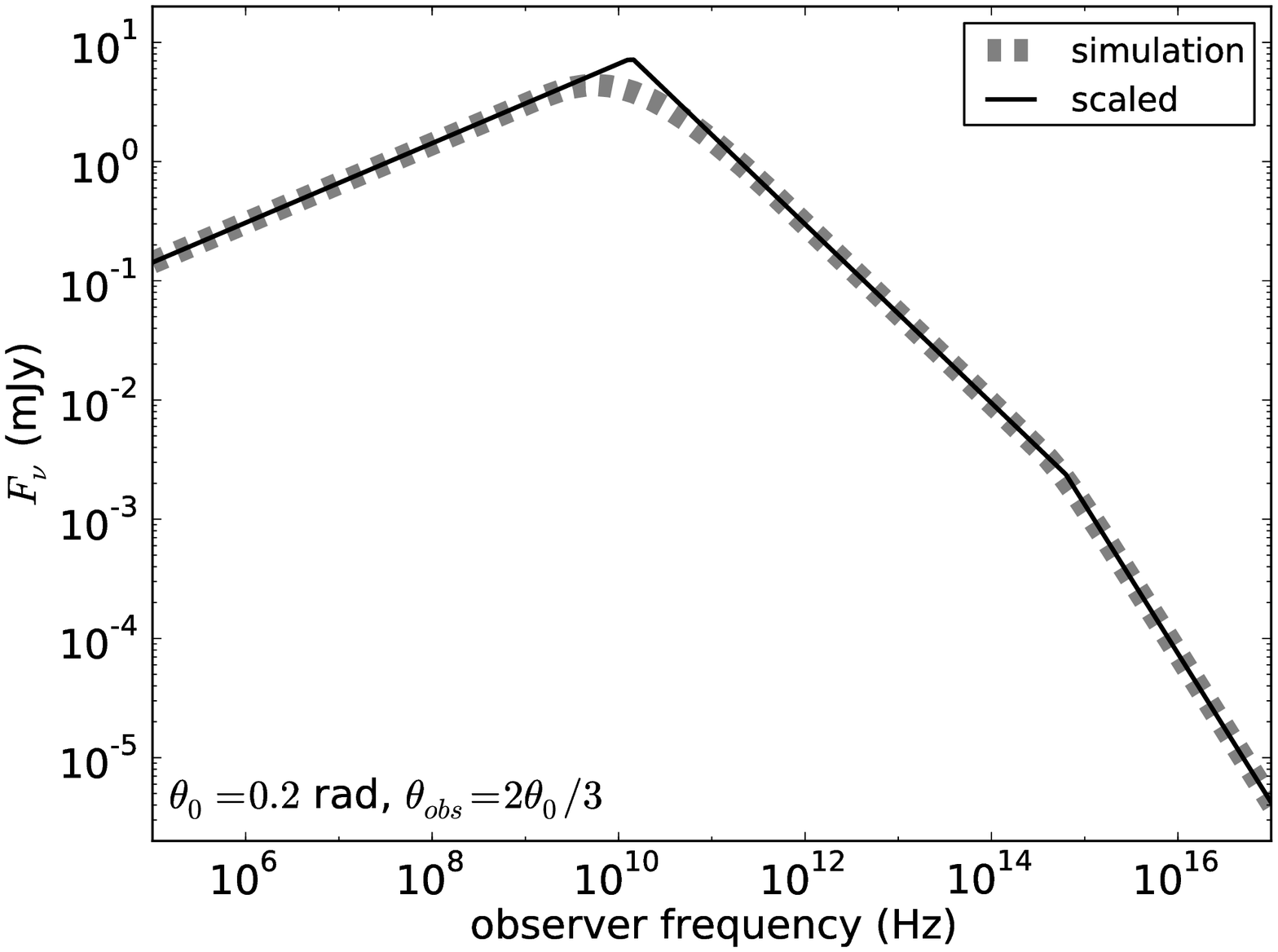}
 \caption{Comparison between simulation and scalings based spectra. The following parameters are used for the spectra: $E_{iso} = 35 \times 10^{52}$ erg, $n_0 = 11 \times 10^{-3}$ cm$^{-3}$, $\theta_0 = 0.2$ rad, $p = 2.5$, $\epsilon_B = 0.03$, $\epsilon_e = 0.1$, $\xi_N = 1.0$, $z = 0$, $d_L = 10^{28}$ cm. The top plot shows a spectrum viewed on-axis, the bottom plot shows the spectrum observed from the average expected observer angle. The spectra are taken at an observer time $t = 43$ days.}
 \label{spectra_figure}
\end{figure}

Using both the on-axis baselines shown in Fig \ref{criticals_figure} and the baselines for $\theta_0 = 0.2$ rad, $\theta_{obs} = 2 \theta_0 / 3$, we have generated spectra (excluding synchrotron self-absorption) for a different set of explosion and radiation parameters and compare these to spectra calculated directly from simulations. The results are shown in Fig \ref{spectra_figure}. The off-axis angle is equal to the average observer angle assuming randomly oriented jets and no detection if $\theta_{obs} > \theta_0$. The scaling approach correctly captures the peak flux and break frequencies. The scalings-based spectra can be further improved upon by including smooth power law transitions between different spectral regimes. Fig \ref{criticals_figure} suggests that some dependency on $\theta_{obs}$ is to be expected.

\section{Summary and Discussion}
\label{discussion_section}

We show that gamma-ray burst afterglow spectra and light curves above the synchrotron self-absorption break can be generated for arbitrary explosion and radiation parameters by scaling the values of a few key parameters ($F_{peak}$, $\nu_c$, $\nu_m$) from a given baseline. The baseline only needs to be calculated once for each observer and jet collimation angle. In the current study we have used sharp transitions between the different spectral regimes, but smooth power law transitions can be parametrized (e.g. following \citealt{Granot2002}). Although we have confined our study to a homogeneous circumburst environment, the generalization to a stellar wind environment is straightforward and will be presented in future work. The blast waves initially follow the ultra-relativistic Blandford-McKee self-similar solution and gradually spread out to the late time non-relativistic Sedov-von Neumann-Taylor stage.

We plot the time evolution of the key parameters $F_{peak}$, $\nu_m$, $\nu_c$ and the plots reveal that the critical frequencies are strongly affected by the jet break. After the jet break, the $\nu_m$ temporal slope quickly drops to the steep late time Sedov-Taylor slope $\nu_m \propto t^{-3}$, while the cooling break $\nu_c$ first steepens then rises to meet the level of its late time asymptote. The steepening of the temporal slope of $\nu_c$ has been observed for GRB 091127 \citep{Filgas2011}, though we caution against overinterpretation of our results given our simplified approach to electron cooling using a global cooling time (following \citealt{Sari1998}). 

The scaling-based light curves fully include all two-dimensional dynamical features of expanding and decelerating afterglow blast waves. The fact that light curves can now be instantly generated for the standard synchrotron afterglow model, while taking realistic dynamics into account, has the potential to strongly impact afterglow light curve fitting. In future work, we will implement the time evolution of key parameters for a wide range of observer and jet collimation angles into a fit code that will be made available for download on \url{http://cosmo.nyu.edu/afterglowlibrary}. Our findings imply that it is now possible in principle to fully fit for the shape of the jet break (e.g. at early time X-ray and optical frequencies, see \citealt{Evans2009, Racusin2009} for examples from \emph{Swift}). More generally, the accuracy of different parametrizations of the jet break transition \citep{Beuermann1999, Harrison1999} can be assessed. With the general shape of the transition into the ST regime being known for arbitrary explosion parameters, the exact moment when the ST asymptote is reached in the light curve (see e.g. \citealt{Livio2000, Wygoda2011}) is no longer relevant, which will improve late time radio calorimetry
(see e.g. \citealt{Berger2004, Shivvers2011}).

In this work we have not discussed synchrotron self-absorption in detail, but we do show that the critical frequency $\nu_a$ obeys the same scaling relations as the other key parameters. The effect of self-absorption will be investigated in a follow-up study. Although the scaling of $\nu_a$ is encouraging from the perspective of broadband afterglow fitting, features such as the chromaticity of the jet-break across the self-absorption break \citep{vanEerten2011jetbreaks} might impact the spectral slope and the sharpness of the transition into the self-absorption spectral regime, rendering a parametrization slightly more complex.

In our previous study \citep{vanEerten2011boxfit} we presented a method to quickly calculate light curves based on scalings on the level of the jet dynamics. This method had the disadvantage that a full radiative transfer was still required for each datapoint. As a result, broadband afterglow fitting-based on this approach still requires use of a large parallel computer in practice. The current method no longer requires radiative transfer calculations and is therefore vastly superior in terms of computational cost. Nevertheless, in two key aspects the `\textsc{box}-based' framework from \cite{vanEerten2011boxfit} remains relevant. First, since the scalings there happen on the level of the dynamics, no parametrizations are necessary in order to describe the transitions between different spectral regimes and smoothly connected power laws emerge naturally. Second, the scalable \textsc{box}-based blast wave dynamics data provide a testing lab for studying the effect of various radiative processes, while the current study takes synchrotron radiation as its starting point.

\acknowledgements

This research was supported in part by NASA through grant NNX10AF62G issued through the Astrophysics Theory Program and by the NSF through grant AST-1009863. The software used in this work was in part developed by the DOE-supported ASCI/Alliance Center for Astrophysical Thermonuclear Flashes at the University of Chicago. We thank Alexander van der Horst and Andrei Gruzinov for helpful comments.\\


\begin{thebibliography}{27}
\expandafter\ifx\csname natexlab\endcsname\relax\def\natexlab#1{#1}\fi

\bibitem[{{Berger} {et~al.}(2004){Berger}, {Kulkarni}, \& {Frail}}]{Berger2004}
{Berger}, E., {Kulkarni}, S.~R., \& {Frail}, D.~A. 2004, \apj, 612, 966

\bibitem[{{Beuermann} {et~al.}(1999){Beuermann}, {Hessman}, {Reinsch},
  {Nicklas}, {Vreeswijk}, {Galama}, {Rol}, {van Paradijs}, {Kouveliotou},
  {Frontera}, {Masetti}, {Palazzi}, \& {Pian}}]{Beuermann1999}
{Beuermann}, K., {Hessman}, F.~V., {Reinsch}, K., {Nicklas}, H., {Vreeswijk},
  P.~M., {Galama}, T.~J., {Rol}, E., {van Paradijs}, J., {Kouveliotou}, C.,
  {Frontera}, F., {Masetti}, N., {Palazzi}, E., \& {Pian}, E. 1999, \aap, 352,
  L26

\bibitem[{{Blandford} \& {McKee}(1976)}]{Blandford1976}
{Blandford}, R.~D., \& {McKee}, C.~F. 1976, Physics of Fluids, 19, 1130

\bibitem[{{Downes} {et~al.}(2002){Downes}, {Duffy}, \&
  {Komissarov}}]{Downes2002}
{Downes}, T.~P., {Duffy}, P., \& {Komissarov}, S.~S. 2002, \mnras, 332, 144

\bibitem[{{Eichler} {et~al.}(1989){Eichler}, {Livio}, {Piran}, \&
  {Schramm}}]{Eichler1989}
{Eichler}, D., {Livio}, M., {Piran}, T., \& {Schramm}, D.~N. 1989, \nat, 340,
  126

\bibitem[{{Evans} {et~al.}(2009){Evans}, {Beardmore}, {Page}, {Osborne},
  {O'Brien}, {Willingale}, {Starling}, {Burrows}, {Godet}, {Vetere}, {Racusin},
  {Goad}, {Wiersema}, {Angelini}, {Capalbi}, {Chincarini}, {Gehrels}, {Kennea},
  {Margutti}, {Morris}, {Mountford}, {Pagani}, {Perri}, {Romano}, \&
  {Tanvir}}]{Evans2009}
{Evans}, P.~A., {Beardmore}, A.~P., {Page}, K.~L., {Osborne}, J.~P., {O'Brien},
  P.~T., {Willingale}, R., {Starling}, R.~L.~C., {Burrows}, D.~N., {Godet}, O.,
  {Vetere}, L., {Racusin}, J., {Goad}, M.~R., {Wiersema}, K., {Angelini}, L.,
  {Capalbi}, M., {Chincarini}, G., {Gehrels}, N., {Kennea}, J.~A., {Margutti},
  R., {Morris}, D.~C., {Mountford}, C.~J., {Pagani}, C., {Perri}, M., {Romano},
  P., \& {Tanvir}, N. 2009, \mnras, 397, 1177

\bibitem[{{Filgas} {et~al.}(2011){Filgas}, {Greiner}, {Schady}, {Kr{\"u}hler},
  {Updike}, {Klose}, {Nardini}, {Kann}, {Rossi}, {Sudilovsky}, {Afonso},
  {Clemens}, {Elliott}, {Nicuesa Guelbenzu}, {Olivares E.}, \&
  {Rau}}]{Filgas2011}
{Filgas}, R., {Greiner}, J., {Schady}, P., {Kr{\"u}hler}, T., {Updike}, A.~C.,
  {Klose}, S., {Nardini}, M., {Kann}, D.~A., {Rossi}, A., {Sudilovsky}, V.,
  {Afonso}, P.~M.~J., {Clemens}, C., {Elliott}, J., {Nicuesa Guelbenzu}, A.,
  {Olivares E.}, F., \& {Rau}, A. 2011, \aap, 535, A57

\bibitem[{{Frail} {et~al.}(2000){Frail}, {Waxman}, \& {Kulkarni}}]{Frail2000}
{Frail}, D.~A., {Waxman}, E., \& {Kulkarni}, S.~R. 2000, \apj, 537, 191

\bibitem[{{Granot} \& {Sari}(2002)}]{Granot2002}
{Granot}, J., \& {Sari}, R. 2002, \apj, 568, 820

\bibitem[{{Granot} \& {Piran}(2011)}]{Granot2011}
{Granot}, J., \& {Piran}, T. 2011, ArXiv e-prints: 1109.6468

\bibitem[{{Harrison} {et~al.}(1999){Harrison}, {Bloom}, {Frail}, {Sari},
  {Kulkarni}, {Djorgovski}, {Axelrod}, {Mould}, {Schmidt}, {Wieringa}, {Wark},
  {Subrahmanyan}, {McConnell}, {McCarthy}, {Schaefer}, {McMahon}, {Markze},
  {Firth}, {Soffitta}, \& {Amati}}]{Harrison1999}
{Harrison}, F.~A., {Bloom}, J.~S., {Frail}, D.~A., {Sari}, R., {Kulkarni},
  S.~R., {Djorgovski}, S.~G., {Axelrod}, T., {Mould}, J., {Schmidt}, B.~P.,
  {Wieringa}, M.~H., {Wark}, R.~M., {Subrahmanyan}, R., {McConnell}, D.,
  {McCarthy}, P.~J., {Schaefer}, B.~E., {McMahon}, R.~G., {Markze}, R.~O.,
  {Firth}, E., {Soffitta}, P., \& {Amati}, L. 1999, \apjl, 523, L121

\bibitem[{{Huang} {et~al.}(1999){Huang}, {Dai}, \& {Lu}}]{Huang1999}
{Huang}, Y.~F., {Dai}, Z.~G., \& {Lu}, T. 1999, \mnras, 309, 513

\bibitem[{{Livio} \& {Waxman}(2000)}]{Livio2000}
{Livio}, M., \& {Waxman}, E. 2000, \apj, 538, 187

\bibitem[{{MacFadyen} \& {Woosley}(1999)}]{MacFadyen1999}
{MacFadyen}, A.~I., \& {Woosley}, S.~E. 1999, \apj, 524, 262

\bibitem[{{Meszaros} \& {Rees}(1997)}]{Meszaros1997}
{Meszaros}, P., \& {Rees}, M.~J. 1997, \apj, 476, 232

\bibitem[{{Paczynski}(1991)}]{Paczynski1991}
{Paczynski}, B. 1991, \actaa, 41, 257

\bibitem[Panaitescu 
\& Kumar(2002)]{PK2002} Panaitescu, A., \& Kumar, P.\ 2002, \apj, 571, 779 

\bibitem[{{Racusin} {et~al.}(2009){Racusin}, {Liang}, {Burrows}, {Falcone},
  {Sakamoto}, {Zhang}, {Zhang}, {Evans}, \& {Osborne}}]{Racusin2009}
{Racusin}, J.~L., {Liang}, E.~W., {Burrows}, D.~N., {Falcone}, A., {Sakamoto},
  T., {Zhang}, B.~B., {Zhang}, B., {Evans}, P., \& {Osborne}, J. 2009, \apj,
  698, 43

\bibitem[{{Rhoads}(1999)}]{Rhoads1999}
{Rhoads}, J.~E. 1999, \apj, 525, 737

\bibitem[{{Sari} {et~al.}(1998){Sari}, {Piran}, \& {Narayan}}]{Sari1998}
{Sari}, R., {Piran}, T., \& {Narayan}, R. 1998, \apjl, 497, L17+

\bibitem[{{Sedov}(1959)}]{Sedov1959}
{Sedov}, L.~I. 1959, {Similarity and Dimensional Methods in Mechanics}
  ({Similarity and Dimensional Methods in Mechanics, New York: Academic Press,
  1959})

\bibitem[{{Shivvers} \& {Berger}(2011)}]{Shivvers2011}
{Shivvers}, I., \& {Berger}, E. 2011, \apj, 734, 58

\bibitem[{{Taylor}(1950)}]{Taylor1950}
{Taylor}, G. 1950, Royal Society of London Proceedings Series A, 201, 159

\bibitem[{{Van Eerten} {et~al.}(2010){van Eerten}, {Leventis}, {Meliani},
  {Wijers}, \& {Keppens}}]{vanEerten2010transrelativistic}
{van Eerten}, H.~J., {Leventis}, K., {Meliani}, Z., {Wijers}, R.~A.~M.~J., \&
  {Keppens}, R. 2010, \mnras, 403, 300

\bibitem[{{Van Eerten} {et~al.}(2010){van Eerten}, {Zhang}, \&
  {MacFadyen}}]{vanEerten2010offaxis}
{van Eerten}, H., {Zhang}, W., \& {MacFadyen}, A. 2010, \apj, 722, 235

\bibitem[{{van Eerten} {et~al.}(2011){van Eerten}, {Meliani}, {Wijers}, \&
  {Keppens}}]{vanEerten2011jetbreaks}
{van Eerten}, H.~J., {Meliani}, Z., {Wijers}, R.~A.~M.~J., \& {Keppens}, R.
  2011, \mnras, 410, 2016

\bibitem[{{Van Eerten} \& {MacFadyen}(2011)}]{vanEerten2011sgrb}
{van Eerten}, H.~J., \& {MacFadyen}, A.~I. 2011, \apjl, 733, L37

\bibitem[{{Van Eerten} \& {MacFadyen}(2011)}]{vanEerten2011jetspreading}
{van Eerten}, H.~J., \& {MacFadyen}, A.~I. 2011, ArXiv e-prints: 1105.2485

\bibitem[{{Van Eerten} {et~al.}(2011){Van Eerten}, {van der Horst}, \&
  {MacFadyen}}]{vanEerten2011boxfit}
{van Eerten}, H.~J., {van der Horst}, A.~J., \& {MacFadyen}, A.~I. 2011, ApJ
  Submitted. ArXiv e-prints: 1110.5089

\bibitem[{{Von Neumann}(1961)}]{vonNeumann1961}
{Von Neumann}, J. 1961, {Collected works}, ed. {von Neumann, J.}

\bibitem[{{Wijers} {et~al.}(1997){Wijers}, {Rees}, \& {Meszaros}}]{Wijers1997}
{Wijers}, R.~A.~M.~J., {Rees}, M.~J., \& {Meszaros}, P. 1997, \mnras, 288, L51

\bibitem[{{Wijers} \& {Galama}(1999)}]{Wijers1999}
{Wijers}, R.~A.~M.~J., \& {Galama}, T.~J. 1999, \apj, 523, 177

\bibitem[{{Woosley}(1993)}]{Woosley1993}
{Woosley}, S.~E. 1993, \apj, 405, 273

\bibitem[{{Wygoda} {et~al.}(2011){Wygoda}, {Waxman}, \& {Frail}}]{Wygoda2011}
{Wygoda}, N., {Waxman}, E., \& {Frail}, D.~A. 2011, \apjl, 738, L23

\bibitem[{{Zhang} \& {MacFadyen}(2006)}]{Zhang2006}
{Zhang}, W., \& {MacFadyen}, A.~I. 2006, \apjs, 164, 255

\bibitem[{{Zhang} \& {MacFadyen}(2009)}]{Zhang2009}
{Zhang}, W., \& {MacFadyen}, A. 2009, \apj, 698, 1261


\end{thebibliography}
\end{document}